\documentclass[%
 twocolumn,
 prb,
 amsmath,amssymb,
 aps,
]{revtex4-1}
\usepackage{float}
\usepackage{graphicx}% Include figure files
\usepackage{dcolumn}% Align table columns on decimal point
\usepackage{bm}% bold math
\usepackage{amsmath}
\usepackage{epstopdf}
\usepackage{textcomp}
\usepackage{color}
\usepackage{lipsum}
\usepackage{ulem}
\usepackage{xcolor}
\usepackage{siunitx}
\begin{document}

\title{Controlled motion of skyrmions in a magnetic antidot lattice}% Force line breaks with \\

\author{J. Feilhauer$^{1}$ } 
\email{Juraj.Feilhauer@savba.sk }
\author{S.~Saha$^{2,3}$}
\email{susmita.saha@psi.ch}
\author{J.~Tobik$^{1}$ } 
\author{M.~Zelent$^{4}$ } 
\author{L. J. Heyderman$^{2,3}$ } 
\author{M.~Mruczkiewicz$^{1}$ }  \email{michal.mruczkiewicz@savba.sk}

\affiliation{%
$^{1}$Institute of Electrical Engineering, Slovak Academy of Sciences, Dubravska Cesta 9, SK-841-04 Bratislava, Slovakia
}%

\affiliation{%
$^{2}$Laboratory for Mesoscopic Systems, Department of Materials, ETH Zurich, 8093 Zurich, Switzerland 
}%
\affiliation{%
$^{3}$Laboratory for Multiscale Materials Experiments, Paul Scherrer Institute, 5232 Villigen PSI, Switzerland
}%

\affiliation{%
$^{4}$Faculty of Physics, Adam Mickiewicz University in Poznan, Umultowska 85, Poznan, PL-61-614 Poland
}%

\date{\today}% It is always \today, today,
       % but any date may be explicitly specified

\begin{abstract}

Future spintronic devices based on skyrmions will require precise control of the skyrmion motion. We show that this goal can be achieved through the use of magnetic antidot arrays. With micromagnetic simulations and semi-analytical calculations based on Thiele equation, we demonstrate that an antidot array can guide the skyrmions in different directions depending on the parameters of the applied current pulse. Despite the fixed direction of the net driving current, due to the non-trivial interplay between the repulsive potential introduced by the antidots, the skyrmion Hall effect and the non-uniform current distribution, full control of skyrmion motion in a 2D lattice can be achieved. 
Moreover, we demonstrate that the direction of skyrmion motion can be controlled by tuning only a single parameter of the current pulse, i.e. current magnitude. For lower current magnitudes the skyrmion can be moved perpendicularly to the current direction, and can overcome the skyrmion Hall effect. For larger current magnitudes, the skyrmion Hall effect can be effectively suppressed and skyrmions can move parallel to the applied current.
\end{abstract}

\pacs{Valid PACS appear here}% PACS, the Physics and Astronomy
\maketitle

\section{\label{sec:intro}Introduction}
In the approximation of continuous magnetization, magnetic skyrmions\cite{Robler_Spontaneus_skyrmion_2006,Fert_skyrmions_on_track_2013,Muhlbauer2009SkyrmionMagnet} are topologically 
non-trivial stable spin textures that can be characterized by a non-zero winding number. 
The winding number for skyrmions in 2D films is $w = \pm 1$, which signifies that the magnetic state
of the skyrmion cannot be unwound into a homogeneous state by a continuous transformation.
Skyrmions in ferromagnetic 
materials are stabilized by the interplay of several magnetic energy contributions, i.e. exchange, dipole, anisotropy and Zeeman. In particular, the 
asymmetric Dzyaloshinskii-Moriya  exchange interaction (DMI)
 introduces chiral canting between neighbouring spins and favours skyrmion stability \cite{boulle2016room,Dzyaloshinsky1958AAntiferromagnetics}.
DMI can arise due to the spin-orbit coupling and the lack of structural inversion symmetry, which can be realized in asymmetric heavy metal/ferromagnet heterostructures\cite{Chen2013NovelFilms}. Due to the interfacial anisotropy, the ferromagnetic layer is magnetized in out-of-plane and the skyrmion is a defect in this homogeneous magnetization with diameter from a few to several hundreds of nanometers. The magnetization of the skyrmion center is oriented in the opposite direction with respect to the rest of the ferromagnetic layer and the magnetization of skyrmion continuously rotates from the center to its edges, so reversing in a radially symmetric fashion.

Another appealing feature of magnetic skyrmions is the possibilty to manipulate them 
by electric current \cite{woo2016observation,Jiang2015BlowingBubbles} via 
spin transfer \cite{koshibae2017theory,Zhang2016MagneticEffect,kim2017current} and spin orbit torques \cite{Litzius2016SkyrmionMicroscopy,litzius2017skyrmion,Reichhardt2015QuantizedSubstrate,yu2016room}.
In general, the current density required to drive the skyrmion motion is much smaller than the current density required to move a domain wall \cite{sampaio2013nucleation,Hrabec2017Current-inducedBilayers, 107}. These unique features of skyrmions make them 
a promising candidates for future spintronic applications such as low dissipation magnetic information storage devices\cite{97}, 
skyrmion racetrack memories\cite{Finocchio2016MagneticApplications}, and logic devices\cite{Zhang2016CreationEngineering}.

%***********************

Unfortunately, skyrmions have a tendency to migrate towards the edges of a magnetic strip, due to the skyrmion Hall effect, which leads to an unstable transverse position and annihilation. Furthermore, the skyrmion motion is  randomized by the 
thermal diffusion and the presence of pining centers in real samples\cite{suess2019spin}. Therefore, key challenges regarding skyrmion motion are related to stabilization, confinement and control of the movement of skyrmions at room temperature \cite{Hirsch1999SpinEffect,Liu2012Spin-TorqueTantalum}.  
The observation that the edges of the sample and defects repel skyrmions led
to the idea to use periodic lattices as a medium with well defined and robust motion of skyrmions\cite{Reichhardt2015QuantizedSubstrate,suess2019spin}.

In order to achieve the control over the skyrmion motion, we consider an antidot lattice realized as a thin ferromagnet/heavy metal bilayer (e.g., Co/Pt, CoFeB/Pt) with circular holes arranged on periodic square lattice.
A hole in the magnetic texture of the ferromagnetic layer (i.e. region with zero magnetization) can be viewed as a
homogeneous layer superposed with a disk with the opposite magnetization orientation. Since the magnetization of this disk is oriented in the same direction as the magnetization of the skyrmion center, the skyrmion is magnetostatically repelled by the antidot.
The repulsion of skyrmion by the antidot lattice can therefore be expressed by a periodic effective potential with energy minima (valleys) located between each four neighbouring antidots.
Due to the dissipation, in the absence of driving torques acting on the skyrmion, the skyrmions are stabilized near the bottom of the valleys. In other words, in the presence of the antidot lattice, the relaxed positions of skyrmions are confined to the discrete square lattice of valleys.
Sufficient spin transfer torque is required to move the skyrmions over the energy barrier from one valley to another.
The torque acting on the skyrmions can be tuned with current pulses of different density and pulse width.
Due to the discrete nature of the valleys, the current pulse parameters required for a skyrmion to arrive at a particular valley create compact regions in the parameter plane encompassing the pulse width, $\Delta t$ and the current density, $j$. Employing a micromagnetic solver Mumax and a semi-analytical model based on the Thiele equation, we calculate a map of these regions in the ($\Delta t$, $j$) plane and study its changes
upon variation of system properties such as damping. 

The skyrmion trajectories in general are not parallel to the current density due to the skyrmion Hall effect.  However, due to the non-trivial interplay between the skyrmion Hall effect and the potential landscape of the antidots, for shorter current pulses with larger magnitudes, the skyrmion can be moved to the valleys in the direction strictly parallel to the net applied current. Thus, for a particular range of current pulse parameters, the skyrmion Hall effect is effectively suppressed. By applying a longer current pulses with smaller magnitudes, the skyrmion can be moved into specific valleys in the direction perpendicular to the current. For this range of current parameters the skyrmion Hall effect is enhanced.

We demonstrate that, for an adequate combination of material, current and antidot lattice parameters, the skyrmion can be transported to almost any nearest-neighbouring valley by applying a specific current pulse. Moreover, transport to the nearest-neighbouring valleys can be achieved by tuning a single parameter of the current pulse, namely the current magnitude.

We therefore propose a method for unprecedented control of the skyrmion placement, utilizing the induced periodic potential of the antidot lattice. This is an important step towards exploiting antidot arrays in future skyrmion based spintronic devices.

The paper is  organized as follows. In Section II, we introduce the numerical and semi-analytical models for skyrmion transport in the antidot lattice driven by an applied current. In Section III, we discuss the results in three parts. To identify some characteristic features of the skyrmion transport in the physical system, we first study a much simpler model with negligible damping and spatially uniform current density (Subsection III A). Then we introduce damping into the model and study the interplay between the potential of the antidot lattice and the skyrmion Hall effect (Subsection III B). Finally, we utilize the gained knowledge to describe the skyrmion transport in the physical system with damping and non-uniform current density resulting from the presence of the antidots (Subsection III C).

\section{\label{sec:intro}Methods}
To study the current driven dynamics of skyrmions in the antidot lattice, we have used the following two methods.
First, we employed the Mumax3 solver to perform micromagnetic simulations, which is a proven method for the investigation 
of magnetization processes. Second, the soliton nature 
of the skyrmion facilitates use of a simpler formalism than that used by the micromagnetic simulations to sufficiently describe the 
skyrmion dynamics, i.e. the Thiele equation. The Thiele equation is solved by numerical time integration. 
In this section we provide a detailed description of the calculations for both approaches.

{\bf Micromagnetic simulations.} 
The governing equation of magnetization dynamics is the Landau-Lifshitz-Gilbert (LLG) equation, which is a partial 
differential equation numerically solved by Mumax3 via a finite difference method. \cite{vansteenkiste2014design}
To simulate the skyrmion transport in the antidot lattice
we have used periodic boundary conditions with a unit cell of square shape with lattice constant of $a=512$~nm and thickness $h=1.4$~nm.
The unit cell contains one antidot realized as a circular hole with diameter $d = 250$~nm.
The rectangular regular discretization mesh has dimensions $2\times2\times1.4$~nm$^3$.
The simulated material was CoFeB with material parameters: saturation magnetization $M_S=1.2\times 10^6$~A/m, DMI constant 
$D=1$~mJ/m$^2$, exchange constant $A_{ex}=1.5\times10^{-11}$~J/m, and damping constant $\alpha$ in the interval from 0.03 to 0.3. To stabilize the skyrmions an out-of-plane magnetic field of 30~mT was applied.

\begin{figure}[!tb] 
%\centering
\centerline{\includegraphics[width=1.0\linewidth]{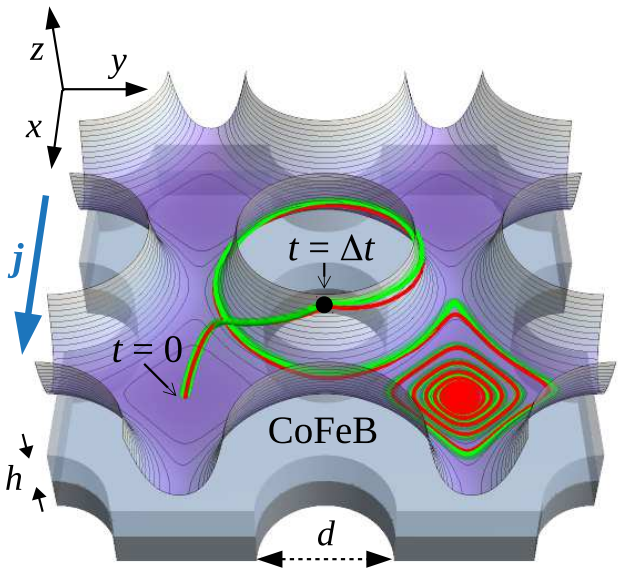}}
\caption{(Color online) 
 Schematic of the ferromagnet/heavy metal heterostructure that hosts the magnetic skyrmions. Antidots are given by circular holes with diameter $d = 250$~nm arranged in the square lattice with lattice constant $a = 512$~nm. The antidots repel the skyrmions as indicated by the potential $V(x,y)$ plotted above the heterostructure. The color curves are an example of a skyrmion trajectory induced by the applied current pulse $\mathbf{j}$ in the $x$-direction. The skyrmion is driven from the relaxed position at the potential minimum (at $t = 0$) to the antidot wall. When the current pulse is switched off (black point at $t = \Delta t$), the skyrmion orbits around the antidot and simultaneously relaxes to the neighbouring potential minimum. The trajectory calculated by Thiele equation (red curve) is in a reasonable agreement with the trajectory calculated micromagnetically (green curve).
}
\label{fig1} 
\end{figure}

In order to describe the motion of the skyrmions induced by the current, we consider the in-plane current flowing directly in the ferromagnetic material. 
For such cases the Zhang-Li form of the  current-induced spin transfer torque is an adequate model \cite{Zhang2004RolesFerromagnets}. We neglect the non-adiabatic contribution to the torque by setting the non-adiabaticity parameter to $\xi=0$.\cite{vansteenkiste2014design} The current densities of rectangular current pulses used in the simulations were varied up to 400~$\mbox{GA/m}^2$, which are the values considered in the experiments on skyrmion motion induced by charge current \cite{woo2018current}. The net current was applied in form of a rectangular pulse in the direction parallel to the row of antidots, along the x-axis (see Fig.\ref{fig1}).

{\bf Thiele equation.}
The soliton character of the skyrmion means that it can be described as a quasiparticle with a fixed magnetization profile. 
This approach considerably reduces the multidimensional spin degrees of freedom, leaving only two parameters - the coordinates of skyrmion center $\mathbf{r}=(x,y)$. The quasiparticle approximation of the skyrmion transforms the LLG equation into the Thiele equation \cite{Thiele1973Steady-StateDomains},
which is an ordinary differential equation describing the dynamics of the skyrmion center located at $\mathbf{r}$.
The translational motion of a skyrmion is affected by the interaction with antidots, which can be represented by a repulsive potential $V(\mathbf{r})$. The motion is driven by the current density $\mathbf{j}(\mathbf{r},t)$ which, for the case of a rectangular current pulse with width $\Delta t$, is given by $\mathbf{j}(\mathbf{r},t) = \mathbf{j}(\mathbf{r}) \Theta(\Delta t - t)$, where $\Theta$ is the Heaviside step function.
All these effects are included in the 
 modified Thiele equation\cite{makhfudz2012inertia,Zhang2016MagneticEffect,Thiele1973Steady-StateDomains}, which takes the form

\begin{equation}
\label{eq:Thiele}
\mathbf{G} \times \left( b_j \mathbf{j}(\mathbf{r},t) - \mathbf{\dot{r}} \right) - \alpha \mathbf{D}  \mathbf{\dot{r}} +\nabla V(\mathbf{r})=0,
\end{equation}
where \textbf{G} is the gyromagnetic coupling vector, $b_j = -\mu_B/e M_S$ ($\mu_B$ is the Bohr magneton and $e>0$ is the electron charge), $\alpha$ is the magnetic damping and $\textbf{D}$ is the dissipative force tensor. 

The first term in (\ref{eq:Thiele}) is the topological Magnus force, perpendicular to the velocity of the skyrmion, which is responsible for the gyrotropic (transverse) motion of the skyrmion \cite{Litzius2016SkyrmionMicroscopy,Jiang2016DirectEffect}. 
In ultrathin films, the magnetization is constant in the $z$ direction and the gyromagnetic coupling vector $\textbf{G}$ is perpendicular to the film plane, i.e. $\textbf{G} = G \textbf{e}_z$ with
\begin{equation}
G = -\frac{4 \pi M_S h w}{\gamma},
\label{Gcko}
\end{equation}
where $\gamma = g \mu_B / \hbar$ is a gyromagnetic ratio and $g$ is an electron g-factor. We assume that the magnetization of the skyrmion center is oriented in the negative $z$-direction. Therefore the winding number of skyrmion used in Eq.(\ref{Gcko}) is $w = -1$.

The second term in (\ref{eq:Thiele}) is the dissipative force, which is linked to the damping in the material and responsible for the friction acting on the skyrmion. In an ultrathin film, the dissipative force tensor $\textbf{D}$ is diagonal and has the form $\textbf{D} = D \delta_{ij}$ with
\begin{equation}
D = -\frac{M_S h \kappa}{\gamma},
\label{Dcko}
\end{equation}
where $\kappa$ is determined by the out-of-plane magnetization profile of the skyrmion. Typically, $\kappa$ is equal to several units of $\pi$  and we use $\kappa = 5.73 \pi$, which provides the best fit of our Thiele equation results with the data obtained by micromagnetic simulations \cite{iwasaki2013universal}.

The third term in (\ref{eq:Thiele}) represents the force acting on the skyrmion due to the repulsive interaction with antidot lattice.
The corresponding periodic potential $V(\mathbf{r})$ was constructed numerically with the use of the micromagnetic solver
in the following manner: 
i) a continuous current was applied to move the skyrmion out of equilibrium to higher energy and the magnetic state was recorded. 
ii) The saved states were used as the initial states for the time evolution with reduced damping constant $\alpha=0.001$ in the absence of current. Each simulation generated quasi-isoenergy contours. 
iii) The generated energy as a function of position was interpolated using a smooth function $V(x,y)$ which has same periodicity as the corresponding antidot lattice.

The resulting periodic repulsive potential $V(x,y)$ is plotted in Fig.~\ref{fig1} with the minimas (valleys) centred between four neighbouring antidots and saddles located between neighbouring antidots.

After algebraic manipulation, Eq.\ref{eq:Thiele} can be written in the form 
\begin{equation}
\dot{\mathbf{r}}(t) = \mathcal{A} \mathbf{j} + \mathcal{B} \mathbf{F}
\label{eq:vec}
\end{equation}
where $\mathbf{F} = - \nabla V$ and
\begin{equation}
\mathcal{A} = \frac{G b_j}{\alpha^2 D^2 +G^2} \begin{bmatrix}
G  &  \alpha D \\
- \alpha D  & G
\end{bmatrix},
\label{mtkyA}
\end{equation}
\begin{equation}
\mathcal{B} = \frac{1}{\alpha^2 D^2  + G^2} \begin{bmatrix}
\alpha D & - G \\
G & \alpha D
\end{bmatrix}.
\label{mtkyB}
\end{equation}
We solved the set of Eqs.(\ref{eq:vec}) numerically with the initial condition $\mathbf{r}(t=0)$ set at the lowest point of the potential landscape $V(\mathbf{r})$, which corresponds to the relaxed position of the skyrmion without an external driving force.

The applicability of the Thiele equation approach was tested by comparing the skyrmion trajectories with calculations performed using the
micromagnetic solver for various values of current density and the pulse width, and we found that the trajectories are in agreement. Examples
of the two trajectories determined with the different methods are shown in the Fig.~\ref{fig1}.  
With the use of Thiele equation, we determined the final position of the skyrmion after the application of a current pulse and 
constructed a map of the final skyrmion positions as the function of the current density and the pulse width.
Applying strong current density might lead to skyrmion anihilation at the disk boundaries.
However, the process of annihilation of the skyrmion by pushing it to an antidot is beyond the validity of the Thiele equation 
approach. Therefore, we have used micromagnetic simulations to determine the critical values of the current pulse parameters, where 
the annihilation starts to occur. 

The antidots realized as holes in the layers do not conduct electric current. The spatially non-uniform current 
density distribution for a geometry with holes was calculated using Comsol.  The non-uniform current density was implemented in the micromagnetic as well as in the Thiele equation calculations. However, the approximation of the uniform current density
simplifies the interpretation of results and is valuable for a qualitative understanding of the magnetic processes. Therefore, in the following we present results for both uniform and non-uniform current distributions.

\section{Results}

To understand the origin of several characteristic features of the skyrmion transport in a real system with damping and non-uniform current distribution, we firstly study much simpler model with negligible damping and uniform current density (Subsection III A). Then we introduce damping into the model and study its effect on the skyrmion trajectory and final position after the current pulse is applied (Subsection III B). Finally, to determine the behaviour of skyrmions in a physical system, we incorporate a non-uniform current distribution into the model and discuss the similarities and differences with the previous simpler cases (Subsection III C). 

\subsection{Damping free system with uniform current}

\begin{figure}[!tb] 
\centering
\includegraphics[width=1.0\linewidth]{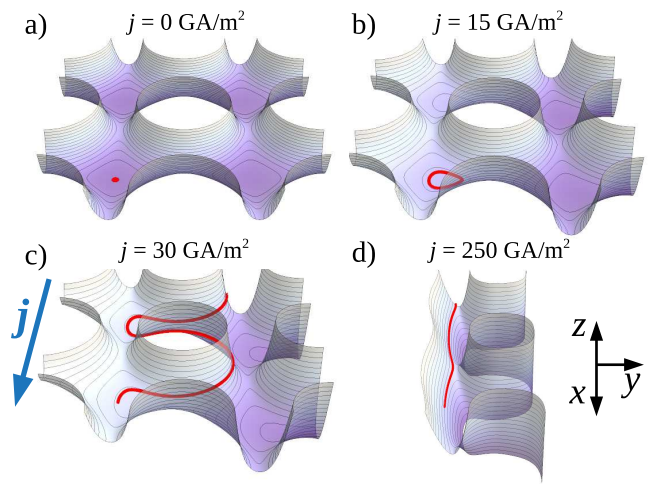}
\caption{(Color online) 
Effective potential $V_{eff}$ combining the effects of antidot potential $V$ and driving current $\mathbf{j}$ on the skyrmion motion in the undamped sample. a)-d) With increasing $\mathbf{j}$ applied in the $x$ direction, the tilt of $V_{eff}$ along the $y$ direction increases. Red curves show the corresponding skyrmion trajectories starting from the potential minimum (valley) of $V$. }
\label{fig2} 
\end{figure}

The main goal of this paper is to study the transport of skyrmions driven by the applied current pulse in the presence of a repulsive antidot lattice. 
For simplicity, we start with the damping free material, i.e. we set $\alpha = 0$. In this case, the matrix $\mathcal{A}$ in (\ref{eq:vec}) is diagonal, which means that the torque acting on the skyrmion is parallel to the current density $\mathbf{j}$ and there is no Hall effect. In contrast, the matrix $\mathcal{B}$ is off-diagonal, which means that the skyrmion is forced to move perpendicularly to the gradient of the antidot potential. Moreover,
if we assume that the current density is spatially uniform, the effect of current can be easily incorporated into the potential via an extra term $V_c = G b_j ( \mathbf{j} \times \mathbf{r} )_z$, simplifying Eq.~\ref{eq:vec} into the form
\begin{equation}
\dot{\mathbf{r}}(t) = (\mathbf{e}_z \times \mathbf{F}_{eff})/G,
\label{eq:Thsimp}
\end{equation}
where $\mathbf{F}_{eff} = -\nabla V_{eff} = -\nabla (V + V_c)$. The dynamics of the skyrmion described by Eq.~\ref{eq:Thsimp} can be readily understood, since the skyrmion moves perpendicularly to the effective force $\mathbf{F}_{eff}$, i.e. it simply follows the isoenergy contour of the effective potential $V_{eff}$.

\begin{figure*}[!tb] 
\centering
\includegraphics[width=0.95\textwidth]{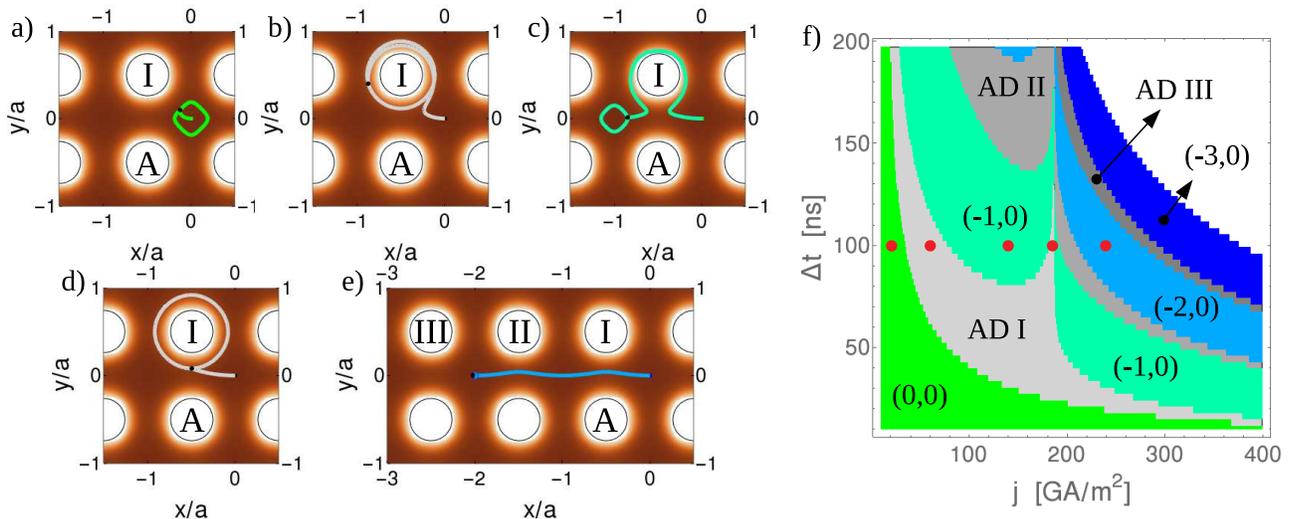}
\caption{(Color online) 
Skyrmion transport without damping driven using rectangular current pulses with the density $j$ and width $\Delta t$. a)-e) Trajectories of skyrmions in the antidot array starting at the bottom of the valley at point $(x,y)=(0,0)$. The value of the pulse width is fixed to $\Delta t = 100$~ns and the the values of $j$ are 20, 60, 140, 187 and 240~$\mbox{GA/m}^2$ respectively. The color of each trajectory corresponds to a particular region in (f). Black dots denote the position of the skyrmion where the current pulse is switched off. f) Map of the final positions of the skyrmion as a function of the current pulse width and density. Red dots indicate the parameters defining the trajectories in (a)-(e).
}
\label{fig3} 
\end{figure*}

When the current is switched off,  $V_{eff} = V$, with closed isoenergy contours centred around the bottom of the valley or around an antidot, depending on whether the corresponding energy is smaller or larger than the energy of the saddle point between the valleys (see Fig.~\ref{fig2}a). When the uniform current is switched on, the resulting effective potential $V_{eff}$ is just the antidot potential $V$ tilted in the direction perpendicular to the direction of the current. The amount of tilt is proportional to the value of the current density, i.e. the larger the current, the larger the tilt of the antidot potential. In particular, when we apply the current in the $x$ direction, the current density has the form $\mathbf{j}(\mathbf{r}) = (j, 0)$ and the effective potential is simplified to
\begin{equation}
V_{eff}(x,y) = V(x,y) + G b_j j y,
\label{Veff}
\end{equation}
i.e. it is equal to the antidot potential $V$ tilted in the $y$-direction. The isoenergy contours of this tilted $V_{eff}$ are no longer just closed but open runaway contours also exist and extend between the valleys in the $x$ direction (see Fig.~\ref{fig2}b-d). 

We now discuss the three types of skyrmion trajectories resulting from the applied current, assuming the skyrmion starts from the bottom of the valley, i.e. the center between four neighbouring antidots (red point in Fig.~\ref{fig2}a). For smaller current density, the corresponding isoenergy contour of $V_{eff}$ is closed and skyrmion oscillates inside the valley (Fig.~\ref{fig2}b). As the current density is increased, the tilt of the antidot potential can be large enough to cause the contour of $V_{eff}$ crossing the starting point to open. In this case, the skyrmion escapes the starting valley and passes along an antidot to the neighbouring valley in the $x$-direction (Fig.~\ref{fig2}c). For sufficiently large current density, the tilt of the antidot potential is so large that the isoenergy contour along which the skyrmion passes to the neighbouring valley can pass directly through the saddle between the valleys avoiding the passage of the skyrmion around the antidot (see Fig.~\ref{fig2}d).

Our aim is to determine, and therefore control, the final positions of the skyrmion after applying a rectangular current pulse with density $j$ and width $\Delta t$. As shown above, when the current is switched on, the skyrmion follows the isoenergy contour of the tilted antidot potential $V_{eff}$. Subsequently, when the current pulse is switched off, the tilted antidot potential returns back to the original antidot potential (without current) and the skyrmion follows a closed isoenergy contour, i.e. it stays trapped inside a valley or orbits around an antidot depending on the parameters $j$ and $\Delta t$. Five examples of such trajectories are shown in Fig.~\ref{fig3}a-e for a skyrmion starting at the valley at $(x,y) = (0,0)$ driven by a current pulse with fixed width $\Delta t = 100 $~ns and various current densities $j$.

The final positions of skyrmion resulting from the applied current pulse are summarized in a pulse width-current density map in Fig.~\ref{fig3}f. This map was calculated using the Thiele equation (\ref{eq:vec}). The regions of parameters $j$ and $\Delta t$ which correspond to the same final position of the skyrmion are given by the same color. The gray regions correspond to the case where the skyrmion orbits around the antidots marked I, II, or III (see e.g. Fig.~\ref{fig3}b). The other colored regions correspond to the situation where the skyrmion ends up trapped inside a valley in the horizontal direction (e.g. at (x,y) =  (0,0), (-1,0), ...). For simplicity, we plot only the data for the valleys with distance smaller than $3a$ from the origin. The data for valleys further than $3a$ are all located in the white region in Fig.~\ref{fig3}b. When the current density is small or the pulse is very short, the skyrmion stays pinned inside the starting valley at (0,0) as depicted by the dark green region. When increasing the pulse parameters, the current pulse is sufficient to move the skyrmion over the saddle point and the skyrmion orbits around the antidot I after the current pulse is turned off (light gray region, AD~I). Increasing the current pulse parameters even more, the skyrmion is able to move to the neighbouring valley (region (-1,0) in light green), orbits around the antidot II (region AD~II in dark gray), moves to the next-neigbouring valley (region (-2,0) in light blue) and so on. There is a discontinuity in Fig.~\ref{fig3}b at $j_u = 187$~GA/m$^2$, which separates two types of trajectories. For $j < j_u$ the skyrmion trajectories crossing the valleys pass around the antidots I, II, ... (e.g. the trajectory for $j = 140$~GA/m$^2$ in Fig.~\ref{fig3}a) while for $j > j_u$ the skyrmions reach these valleys directly through the saddles of the antidot potential (e.g. the trajectory for $j = 240$~GA/m$^2$ in Fig.~\ref{fig3}a). At $j = j_u$ the skyrmion passes to the saddle point of the effective potential $V_{eff}$ located between the antidots I and A where $\bold{F}_{eff} = \nabla V_{eff} = 0$. Here, the torque acting on the skyrmion due to the applied current is completely compensated by the repulsion of the antidot lattice. Then, since the right side of the Thiele equation (\ref{eq:Thsimp}) vanishes, the skyrmion velocity becomes zero and it stops, i.e. the saddle point of $V_{eff}$ is an unstable position for the skyrmion. After the current pulse is switched off the skyrmion starts to orbit around the antidot I. The corresponding skyrmion trajectory is shown in Fig.~\ref{fig3}d where the black point indicates the unstable position of the skyrmion where the skyrmion is fixed until the current pulse is switched off.

\begin{figure}[!tb] 
\centering
\includegraphics[width=0.95\linewidth]{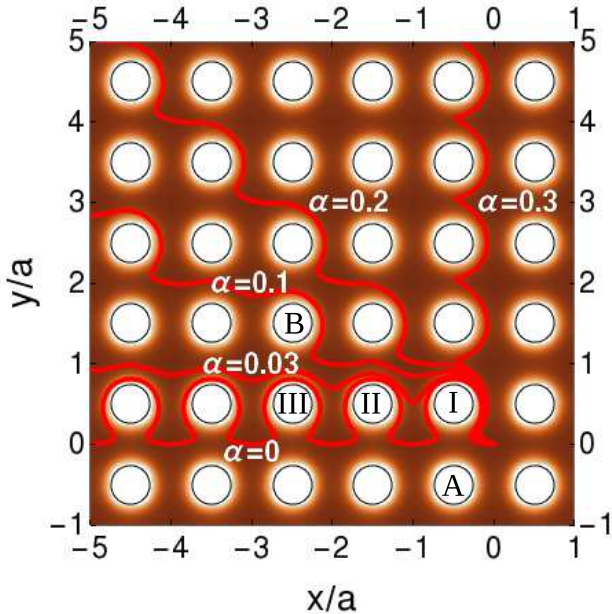}
\caption{(Color online) 
Trajectories of a skyrmion starting in the valley at (0,0) for various values of damping. The skyrmion motion is driven by the uniform current density $j = 90$~GA/m$^2$ applied along the $x$ direction. In the undamped case ($\alpha = 0$) the trajectory extends only to the valleys in the direction of the current (e.g. at (-1,0), (-2,0), ...) while, for non-zero damping, the trajectories also reach the valleys in the transverse direction (e.g. at (-1,1), (0,1), ...).}
\label{fig4} 
\end{figure}

As shown above, without damping, it would be possible to transfer the skyrmions only to neighbouring valleys in the direction parallel with the current. As we show in the following section, to move the skyrmion in the transverse direction with respect to the current, nonzero damping and the resulting skyrmion Hall effect is essential.

\subsection{Damped system with uniform current}
In order to determine how the motion of the skyrmion is affected by damping, we include a non-zero $\alpha$ in the off-diagonal elements of matrix $\mathcal{A}$ in Eq.~\ref{eq:vec}, which is responsible for the skyrmion Hall effect. In the unpatterned thin film, the skyrmion would move at an angle with respect to the current direction given by a Hall angle
\begin{equation}
\label{eq:Hall}
|\Theta_{H}|=\arctan{\frac{\kappa \alpha}{4 \pi}}.
\end{equation}
From Eq.~\ref{eq:Hall}, we found that the value of $\Theta_{H}$ varies from  $\ang{2.5}$ to $\ang{23.3}$ when changing $\alpha$ from $0.03$ to $0.3$. Another effect resulting from the damping is the relaxation of the skyrmion to a lower energy configuration when the current pulse is switched off. Due to the damping, the diagonal elements of $\mathcal{B}$ in Eq.~\ref{eq:vec} are nonzero and the skyrmion motion converges to the local energy minima, i.e. it ends up at the bottom of a valley.

Due to the nontrivial manifestation of damping in the Thiele equation (\ref{Veff}), the simple concept of a tilted antidot potential described by (\ref{eq:Thsimp}) is not valid for non-zero damping. Therefore the shapes of the skyrmion trajectories are more complex, which is illustrated in Fig.~\ref{fig4} for fixed uniform current density $j = 90$~GA/m$^2$ and various values of damping $\alpha$. In the undamped case ($\alpha = 0$), the trajectory extends in the $x$ direction with a periodic form with a period of a single lattice constant $a$. This is a result of the fact that the direction of the applied current and the corresponding torque acting on the skyrmion are parallel to the symmetry axis of the antidot lattice (i.e. $x$ axis). If the torque applied on the skyrmion had a nonzero component in the $y$ direction, the skyrmion trajectory would on average follow the torque direction but it would be formed from discrete steps with the periodicity larger than a single $a$. For non-zero damping, since the current flows along the $x$ direction, the effective $y$ component of the torque is generated via the skyrmion Hall effect. Therefore, as was also shown in Ref.~[\onlinecite{Reichhardt2015QuantizedSubstrate}], depending on the amount of damping and density of the applied current, the skyrmion trajectories in the presence of an antidot lattice form a series of periodic discrete steps extending in both $x$ and $y$ directions with periods that can be multiples of $a$, e.g. the trajectory for $\alpha = 0.1$ in Fig.~\ref{fig4}.  With the increase of the damping, the angle between the skyrmion trajectory and the current direction along the $x$ axis also increases. Due to the presence of a periodic antidot potential, the trajectory angle is much larger than the corresponding Hall angle for the unpatterned film. For example, the trajectory for $\alpha = 0.3$ is perpendicular to the direction of the current while the corresponding Hall angle is only $\ang{23.3}$. Therefore, the presence of the antidot lattice enhances the skyrmion Hall effect. Similar enhancement of skyrmion Hall effect was observed in Ref.\onlinecite{Reichhardt2015QuantizedSubstrate} for a skyrmion moving on a two-dimensional periodic substrate.

As we have seen, the skyrmion trajectories in a damped system extend not only to the valleys in the direction of the current but also to the valleys in the transverse direction. Therefore, the number of valleys that are reachable by the skyrmion is significantly increased. When considering a finite current pulse, we can distinguish two scenarios for the skyrmion transfer between the neighbouring valleys. First, the skyrmion is directly transferred to the desired valley by an applied current and then, when the current is switched off, the skyrmion relaxes to the valley bottom. 
The second case is illustrated in Fig.~\ref{fig1} where, after the current pulse is switched on, the skyrmion climbs from the bottom of the valley to the antidot wall until the pulse is switched off (black point). Without damping the skyrmion would orbit around the antidot forever but, since damping is present ($\alpha = 0.03$), the skyrmion loses its energy and relaxes in a spiral trajectory to the bottom of one of the four valleys next to the antidot. 

 \begin{figure}[!tb] 
\centering
\includegraphics[width=1.0\linewidth]{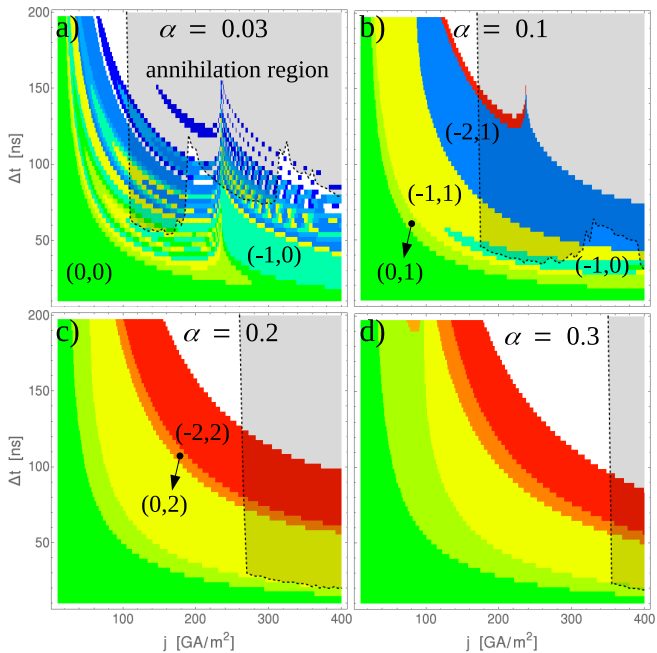}
\caption{(Color online) 
Final position of the skyrmion after the application of uniform current pulse with density $j$ and width $\Delta t$. Color regions correspond to the parameters with the same final valley. The damping is a) $\alpha = 0.03$, b) $\alpha = 0.1$, c) $\alpha = 0.2$ and d) $\alpha = 0.3$. Shaded region calculated by micromagnetic solver shows the parameters of current pulse at which the skyrmion is annihilated.}
\label{fig5} 
\end{figure}

As for the undamped case in Fig.~\ref{fig3}b, in Fig.~\ref{fig5} we show the final positions of the skyrmion that result after applying a uniform current pulse with density $j$ and width $\Delta t$. The maps in Fig.~\ref{fig5} were calculated using the Thiele equation (\ref{eq:vec}) for various values of damping constant $\alpha$. 
%The regions of parameters $j$ and $\Delta t$ which correspond to the same final position of the skyrmion (i.e. the same valley) are marked by the same color. For simplicity, we plot only the data for the valleys with distance smaller than $3a$ from the origin. The data for valleys further than $3a$ are all included in the white region.
For small damping, $\alpha = 0.03$ (Fig.~\ref{fig5}a), we can identify several features that also appear in the undamped case. When the current density or width of the current pulse is too small, the skyrmion just stays pinned in the starting valley (dark green region) after the pulse is applied. The current $j_u$ defining the trajectories passing to the unstable position of the skyrmion located between the antidots I and A is shifted to the larger values, i.e. $j_u = 234$~GA/m$^2$.
\begin{figure*}[!t] 
\centering
\includegraphics[width=0.65\textwidth]{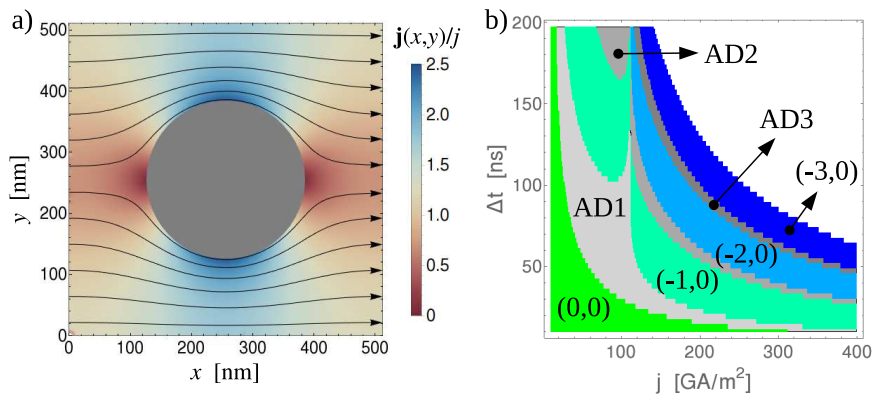}
\caption{(Color online) 
a) Distribution of a periodic non-uniform current density $\bold{j}(x,y)$ in a single unit cell of the antidot lattice. $\bold{j}(x,y)$ was calculated using the Comsol package where the antidot hole (gray circle) is assumed to be nonconductive. b) Map of the final positions of the skyrmion after application of current pulses with non-uniform current density in the undamped system.}
\label{fig6} 
\end{figure*}
For $j > j_u$, the region corresponding to the transport of skyrmion to the neighbouring valley (-1,0) directly through the saddle between the antidots I and A has almost the same area and shape as in the undamped case. However, due to the enhanced skyrmion Hall effect, the same valley is almost unreachable via the trajectories passing around the antidot I and the region (-1,0) is much smaller for $j < j_u$ than for the undamped case. The former light gray region ascribed to the trajectories orbiting around the antidot I (region AD~I in Fig.~\ref{fig3}f) is now decomposed into the set of thin regions corresponding to the four valleys at $(x,y) = (0,0)$, (-1,0), (-1,1) and (0,1) next to the antidot I. As discussed above, due to the damping, the skyrmion orbiting around the antidot I looses its energy and ends up in one of the four neighbouring valleys. When the damping is small, the skyrmion can encircle the antidot several times until it relaxes to the final valley. Therefore, the final valley is strongly sensitive to the position of the skyrmion on the antidot wall at the moment when the current pulse is swiched off. Similarly, the other gray regions from Fig.~\ref{fig3}c describing the orbiting skyrmion around the antidots II, III, ... decompose into the regions corresponding to the surrounding valleys. This is the reason for rather chaotic structure of the map Fig.~\ref{fig5}a for larger values of parameters $j$ and $\Delta t$. 

For larger damping $\alpha = 0.1$ (Fig.~\ref{fig5}b) the value of $j_u = 430$~GA/m$^2$ exceeds the displayed interval of current densities. Therefore, in Fig.~\ref{fig5}b all skyrmions traveling along the trajectories which leave the starting valley pass around the antidot I. Then the only way that the skyrmion can reach the valley (-1,0) is by descending from the wall of antidot I after the current pulse is off (light green region in Fig.~\ref{fig5}b). Since the potential of the antidots $V(x,y)$ and the applied electric current $\mathbf{j}$ have periodicity of the antidot lattice, the points related to the unstable position of the skyrmion are periodically localized throughout the whole antidot lattice. As we have already discussed, the periodicity of skyrmion trajectories can exceed a single lattice constant, which means that a skyrmion can pass to an unstable point several lattice constants away from the starting point. As an example, the discontinuity appearing in the blue region at (-2,1) for $j = 240$~GA/m$^2$ corresponds to the trajectory that passes to the unstable point located between the antidots III and B defined in Fig.~\ref{fig4}. 

For even larger values of damping, the valley (-1,0) is unreachable within the given range of current densities as is demonstrated in Fig.~\ref{fig5}c,d for $\alpha = 0.2$ and $\alpha = 0.3$, respectively.
To summarize, for $\alpha = 0.1$, the regions in the parameter plane $(\Delta t,j)$ corresponding to the different final positions of the skyrmion are more compact than for the case of small damping $\alpha = 0.03$ where the situation is quite chaotic. In contrast, for larger values of damping $\alpha = 0.2, 0.3$, the region corresponding to the valley (-1,0) is shifted to the large values of current outside of the displayed interval. Therefore, the number of neighbouring valleys the skyrmion can be moved to is reduced for $\alpha = 0.2, 0.3$. Therefore, to maximize the number of reachable valleys by the skyrmion, the value of damping $\alpha = 0.1$ is optimal for the given parameters of the antidot lattice and material properties.

For larger values of $j$ or $\Delta t$, the skyrmion inelastically scatters off the antidots or even annihilates at the antidot edges. The skyrmion annihilation is not captured by the Thiele equation since it is based on the assumption that the magnetization profile of skyrmion is fixed. The transparent gray regions in Fig.~\ref{fig5} highlight the parameters of current pulses which lead to the skyrmion annihilation calculated using micromagnetic simulations. For smaller values of damping, the regions of annihilation cover a significant part of the displayed parameter maps. 

 \begin{figure}[!tb] 
\centering
\includegraphics[width=1.0\linewidth]{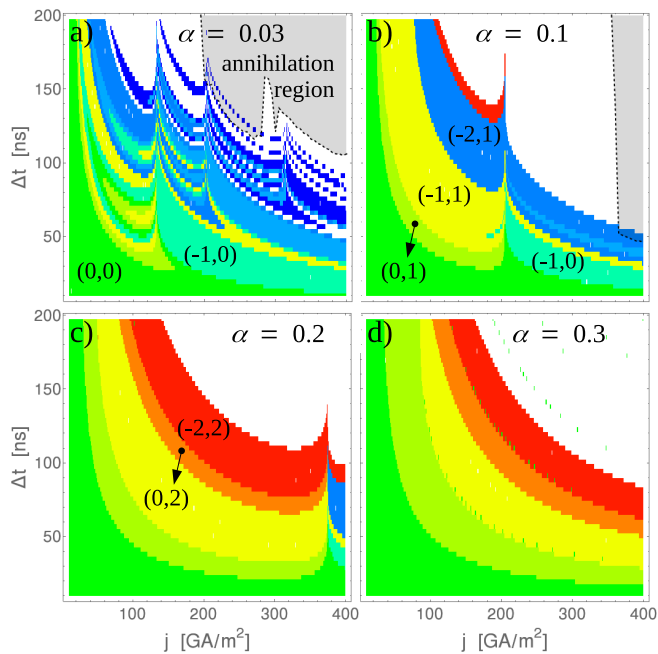}
\caption{(Color online) 
Final position of the skyrmion after the application of non-uniform current pulse with density $j$ and width $\Delta t$. Color regions correspond to the parameters with the same final valley. The coordinate system is shown in the Fig.~ 4. The damping is a) $\alpha = 0.03$, b) $\alpha = 0.1$, c) $\alpha = 0.2$ and d) $\alpha = 0.3$. Shaded region calculated by micromagnetic solver shows the parameters of current pulse at which the skyrmion is annihilated.}
\label{fig7} 
\end{figure}

\subsection{System with non-uniform current}

Here we employ a more realistic model of electric current density. Up till now we have assumed that the electric current density is distributed uniformly and flows along the $x$-direction in the whole antidot lattice, i.e. $\mathbf{j}(x,y) = (j,0)$. However, since the antidot is a hole in the heterostructure, the current cannot enter the antidot and has to flow around it. Therefore, using the Comsol software package, we have calculated a more realistic non-uniform current density $\mathbf{j}(x,y)$ assuming a net voltage drop applied in the $x$ direction. This periodic non-uniform current distribution $\mathbf{j}(x,y)$ is shown in Fig.~\ref{fig6}a. Due to the symmetry of the square antidot lattice, the current density averaged along the transversal ($y$) direction is the same as in the uniform case, i.e. $\frac{1}{a}\int_{0}^{a} \mathbf{j}(x,y) dy = (j,0)$ and, as a result of current conservation, it is independent of $x$. We have incorporated this non-uniform current distribution into the micromagnetic simulation and Thiele equation (\ref{eq:vec}) and recalculated the maps of final skyrmion positions in the current pulse parameter plane. 

The map for non-uniform current density and zero damping is shown in Fig.~\ref{fig6}b. This map is very similar to the case of uniform current in Fig.~\ref{fig3}b, but the discontinuity corresponding to the unstable position of skyrmion is shifted to smaller values of current. 

The skyrmion transport driven by short current pulses with non-uniform current density is summarized in Fig.~\ref{fig7} for various values of damping. Again, these maps are similar to the maps for uniform current in Fig.~\ref{fig5}, but there are some significant differences. In general, as a result of the current flowing around the antidot, the currents needed to move the skyrmion directly through the saddle to the neighbouring valley (avoiding the orbiting around an antidot) are smaller. For small damping, $\alpha = 0.03$, there are multiple discontinuities corresponding to the unstable positions of skyrmion located between the antidots along the $x$ direction from the starting valley. The light green region corresponding to the valley (-1,0) is broader, which makes this valley accessible for a larger range of current pulse parameters. Most importantly, the micromagnetic simulations reveal that the annihilation regions are shifted to the larger values of current pulse parameters than for uniform current in Fig.~\ref{fig5}. This is expected since, for uniform current, the current flow at one side of the antidot pushes the skyrmion directly towards the antidot edge, while for non-uniform current, the current flowing around the antidot reduces the contact of the skyrmion with the antidot edge. For $\alpha = 0.1$ (Fig.~\ref{fig7}b), the annihilation does not affect the regions corresponding to the nearest-neighbouring valleys, which makes this map suitable for the well-defined and robust control of the skyrmion motion. 

\begin{figure}[!tb] 
\centering
\includegraphics[width=0.95\linewidth]{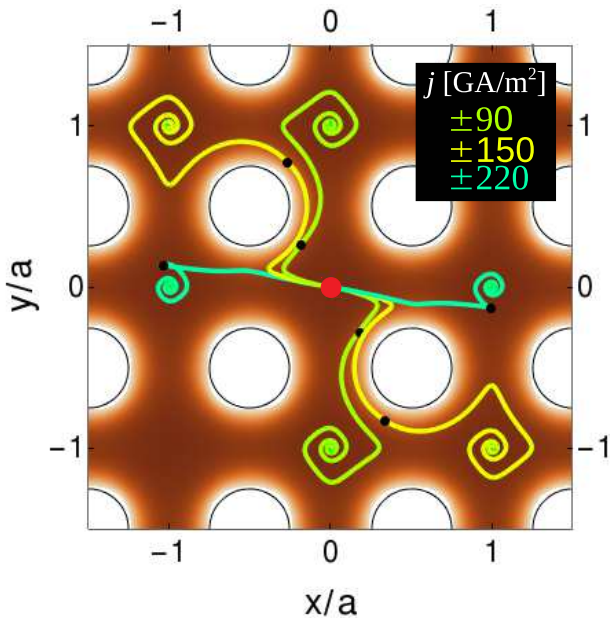}
\caption{(Color online) 
Skyrmion trajectories of damped system with $\alpha = 0.1$, non-uniform current density, fixed width of current pulse at $\Delta t = 50$~ns and varying current densities and polarities. By adjusting a single parameter (j) of the current pulse, the skyrmion can be transfered to almost all of the neighbouring valleys in the longitudinal as well as transversal direction. 
}
\label{fig8} 
\end{figure}

To demonstrate the controlled motion of skyrmions in the antidot lattice we present the response of the skyrmion driven by current pulses applied in the $x$ direction with fixed width $\Delta t = 100$~ns and varying density and polarity. The skyrmion trajectories corresponding to $j = \pm 90, \pm 150, \pm 220$~GA/m$^2$ are shown in Fig.~\ref{fig8}. By tuning a single parameter, current density $j$, the skyrmion starting at the bottom of a valley can be transported to six of the eight neighbouring valleys. 

\section{Summary}

In this work, we have developed a semi-analytical model to determine the skyrmion motion driven by short in-plane current pulses in the presence of a magnetic antidot array. Due to the repulsion of the skyrmion by the antidot edges, the antidot lattice results in an effective potential of attractive valleys located between each four neighboring antidots. We have demonstrated that skyrmion transport between individual valleys can be controlled by applying a rectangular current pulse with adequate density and width. As a result of the interplay between the antidot potential, skyrmion Hall effect and non-uniformity of the current, skyrmions can be manipulated in the longitudinal and even in the transverse direction with respect to the current flow. We have identified two mechanisms determining the final position of the skyrmion: i) the skyrmion is directly driven by the applied current to the desired valley, ii) after the current pulse is switched off, the skyrmion relaxes down the antidot wall to the desired valley. 

We have calculated maps showing the regions of the current pulse parameters that give a particular final position of the skyrmion after the pulse is switched off. Starting from the bottom of a valley, our calculations show that, by applying an adequate unidirectional current pulse, it is possible to move the skyrmion to almost all of the neighboring valleys horizontally and vertically. Therefore, by using a sequence of electrical current pulses, the magnetic antidot arrays can be used as a medium for well controlled skyrmion motion. Our results are therefore an important step towards skyrmion based devices.

\section{Acknowledgments}
 We acknowledge funding from the Slovak Grant Agency APVV, grant number APVV-16-0068 (NanoSky), JF acknowledges the project VEGA 2/0162/18 and JT acknowledges the VEGA project 2/0150/18, MM acknowledge funding from the EU FP7 SASPRO Programme (REA Grant Agreement No. 609427, project WEST 1244/02/01) and further co-funded by the Slovak Academy of Sciences. SS acknowledges ETH Zurich Post-Doctoral fellowship and Marie Curie actions for People COFUND program (Grant No. FEL-11 16-1). MZ acknowledges funding from the Adam Mickiewicz University Foundation, National Science Centre, Poland Grant No. UMO-2017/27/N/ST3/00419 and National Scholarship Programme of the Slovak Republic.

%\bibliographystyle{apsrev4-1}
%\bibliography{feilhauer.bib}

\end{document}